\documentclass[aip,reprint]{revtex4-1}

\usepackage[margin=.75in]{geometry}

\usepackage{amsmath,amssymb}

\usepackage{color}
\definecolor{indigo}{RGB}{0,0,120}
\usepackage[colorlinks=true, linkcolor=indigo, citecolor=blue, urlcolor=indigo]{hyperref}

\newcommand{\tl}[1]{\tilde{#1}}
\newcommand{\dd}[2]{\frac {\partial #1}{\partial #2}}
\newcommand{\pdr}{\partial}

\newcommand{\grad}{{\bf \nabla}}

\newcommand{\beq}{\begin{equation}}
\newcommand{\eeq}{\end{equation}}
\newcommand{\beqs}{\begin{eqnarray}}
\newcommand{\eeqs}{\end{eqnarray}}

\newcommand{\half}{\frac{1}{2}}
\newcommand{\ov}[1]{\frac{1}{#1}}

\def\al{\alpha} 		
\def\del{\delta}
\def\g{\gamma} 
\def\eps{\epsilon} 
\def\la{\lambda}
\def\sig{\sigma}
\def\om{\omega}		

\newcommand{\bfS}{{\bf S}}
\newcommand{\bff}{{\bf f}}

\newcommand{\bfv}{{\bf v}}
\newcommand{\bfw}{{\bf w}}
\newcommand{\bfy}{{\bf y}}
\newcommand{\bfj}{{\bf j}}
\newcommand{\bfr}{{\bf r}}
\newcommand{\bfA}{{\bf A}}
\newcommand{\bfE}{{\bf E}}
\newcommand{\bfB}{{\bf B}}
\newcommand{\bfT}{{\bf T}}

\begin{document}


\title{Conservative regularization of compressible dissipationless two-fluid plasmas}

\author{Govind S. Krishnaswami} 

\affiliation{Chennai Mathematical Institute,  SIPCOT IT Park, Siruseri 603103, India}

\email{govind@cmi.ac.in, sonakshi@cmi.ac.in}

\author{Sonakshi Sachdev} 

\affiliation{Chennai Mathematical Institute,  SIPCOT IT Park, Siruseri 603103, India}

\author{Anantanarayanan Thyagaraja}

\affiliation{Astrophysics Group, University of Bristol, Bristol, BS8 1TL, UK}

\email{athyagaraja@gmail.com}

\date{28 Feb, 2018. \quad Published in Physics of Plasmas {\bf 25}, 022306 (2018). arXiv:1711.05236v2}

\begin{abstract}

This paper extends our earlier approach [cf. Phys. Plasmas {\bf 17}, 032503 (2010), {\bf 23}, 022308 (2016)] to obtaining \`a priori bounds on enstrophy in neutral fluids (R-Euler) and ideal magnetohydrodynamics (R-MHD). This results in a far-reaching local, three-dimensional, non-linear, dispersive generalization of a KdV-type regularization to compressible/incompressible dissipationless two-fluid plasmas and models derived therefrom (quasi-neutral, Hall and ideal MHD). It involves the introduction of vortical and magnetic `twirl' terms $\lambda_l^2 ({\bf w}_l + \frac{q_l}{m_l} {\bf B}) \times (\nabla \times {\bf w}_l)$ in the ion/electron velocity equations ($l = i,e$) where ${\bf w}_l = \nabla \times {\bf v}_l$ are vorticities. The cut-off lengths $\lambda_l$ must be inversely proportional to the square-roots of the number densities $(\lambda_l^2 n_l = C_l)$ and may be taken as Debye lengths or skin-depths. A novel feature is that the `flow' current $\sum_l q_l n_l {\bf v}_l$ in Amp\`ere's law is augmented by a solenoidal `twirl' current $\sum_l \nabla \times \nabla \times \lambda_l^2 {\bf j}_{{\rm flow},l}$. The resulting equations imply conserved linear and angular momenta and a positive definite swirl energy density ${\cal E}^*$ which includes an enstrophic contribution $\sum_l (1/2) \lambda_l^2 \rho_l {\bf w}_l^2$. It is shown that the equations admit a Hamiltonian-Poisson bracket formulation. Furthermore, singularities in $\nabla \times {\bf B}$ are conservatively regularized by adding $(\lambda_B^2/2 \mu_0) (\nabla \times {\bf B})^2$ to ${\cal E}^*$. Finally, it is proved that among regularizations that admit a Hamiltonian formulation and preserve the continuity equations along with the symmetries of the ideal model, the twirl term is unique and minimal in non-linearity and space derivatives of velocities.

\end{abstract}

\maketitle
\section{Introduction}
\label{s:intro}
Plasma physics finds extensive applications in astrophysics, physics of fusion devices like tokamaks, stellarators and in inertial confinement and in technological applications\cite{Hazeltine-Meiss, Wesson, Lifshitz-Pitaevski, Rosenbluth-Sagdeev, Kulsrud, Arnab, Michel}. Plasmas have extremely complex dynamics when they interact with self-generated and externally applied electric and magnetic fields. The dynamics of such systems are governed both by Maxwell's equations and either a kinetic or fluid model representing the co-evolution of the plasma variables. In kinetic descriptions appropriate distribution functions are introduced for the ions and electrons of the plasma. They are evolved according to equations such as the Boltzmann-Fokker-Planck system. The charge and current densities derived from the distribution functions are then used to evolve the fields. In fluid models only the first few ``principal moments'' like the number densities, velocities, temperatures, stresses and heat fluxes appear. It is often the case that the fluid description provides a relatively tractable system which can be used to describe a variety of phenomena actually observed in experiments and in the cosmos. Among fluid models, the simplest ones are generalisations of the well-known dissipationless Euler equations of neutral fluid dynamics to include the effects of electromagnetic body forces. A typical example is provided by the classic model known as Ideal Magneto Hydrodynamics [``ideal MHD'', see, for example,\cite{Goedbloed-Poedts, JBT}] which has found very wide application in both fusion plasma theory and in astrophysical theories. This theory was used by Alfv\'{e}n to describe plasma waves in a magnetised fluid [see the classic text by Stix \cite{Stix}] and to show that in the absence of dissipation [resistivity and viscosity and possibly thermal diffusivity] the magnetic field is ``frozen'' into the flow. This result has wide application to both solar physics and to important classes of instabilities known to occur in tokamak plasmas [``ideal ballooning and kink modes'' {\it op.cit} \cite{Hazeltine-Meiss,Wesson,Goedbloed-Poedts, JBT}].

It is generally the case that even the simplest ideal MHD description involves rather complicated nonlinear partial differential equations. One does not have useful exact, analytically derived solutions valid for experimentally relevant situations. The only generally applicable methods are numerical methods. The dissipationless two-fluid (ion and electron) equations are similar in their qualitative properties to the Euler equations of inviscid fluid dynamics and ideal MHD. They possess several conservation laws but involve energy transfer mechanisms which can lead to short-wavelength singularities like vortex and current sheets, shocks and finite-time unbounded behaviour of mean-square vorticity(``enstrophy'') and current density. It is usually the case that ``ultra-violet'' singularities of these types are resolved by viscosity, thermal conductivity and electrical resistivity. All these are entropy-producing effects and are not consistent with the conservation properties of the dissipationless models. Numerical solutions of the conservative equations can become singular when evolved. It is important to distinguish between purely numerical instabilities which have nothing to do with physical properties of the system and real physical instabilities. For these reasons, it is useful to extend methods developed in our earlier work to `regularize' the Euler and ideal MHD models to two-fluid plasma models. In this work we describe this extension which also has more fundamental applications to the formulation of statistical theories of the dynamics of the systems considered.

In \cite{thyagaraja,govind-sonakshi-thyagaraja-pop,govind-sonakshi-thyagaraja-arXiv} new conservative regularizations of incompressible and compressible Eulerian flow and ideal MHD were introduced. These are three-dimensional nonlinear dispersive but dissipationless counterparts of the Navier-Stokes and visco-resistive MHD equations, just as the KdV equation is a dispersive but inviscid counterpart of the one-dimensional viscous Burgers equation \cite{whitham}. The primary motivation was to regulate possible vortical singularities by ensuring an \`a priori bound on enstrophy. The guiding principles in the choice of regularizing terms were that they be local, minimal in non-linearity and derivatives of velocity $\bfv$, small enough to leave macro-scale dynamics unchanged and preserve Galilean, parity and time-reversal symmetries of the ideal equations. These principles led us to regularized models called R-Euler and R-MHD that involved a new `twirl' term $\la^2 \: \bfw \times (\grad \times \bfw)$ in the velocity equation (i.e., Newton's law) and a corresponding term $\grad \times (\la^2 (\grad \times \bfw) \times \bfB)$ in Faraday's law. These terms correspond to the addition of a  vortical energy density $(1/2) \la^2 \rho \; \bfw^2$ to the flow energy density $(1/2)\rho \bfv^2$. Here $\bfw = \grad \times \bfv$ is the vorticity while $\rho$ is the mass density satisfying the continuity equation. The regulator $\la$ acts as a short-distance cut-off to the growth of enstrophy and must satisfy the constitutive law $\la^2 \rho = $ constant for a conserved energy to exist. Thus $\la$ is like a position-dependent mean free path: smaller in denser regions. Like viscosity $(\nu \grad^2 \bfv)$, the twirl term is second order in velocity derivatives, but unlike the former, it is non-linear and non-dissipative. Indeed, the equations were shown to admit a Hamiltonian-Poisson bracket formulation and local conservation laws for energy, linear and angular momenta, flow/magnetic and cross helicities. Regularized analogues of the Kelvin-Helmholtz and Alfv\'en theorems were obtained, demonstrating that vorticity and magnetic field are frozen into a swirl velocity $\bfv_* = \bfv + \la^2 \: \grad \times \bfw$. 

In Section \ref{s:reg-eqns-2-fluid-compress}, we extend our local conservative regularization of compressible ideal MHD to non-relativistic two fluid (ion-electron) plasmas. The extension to multi-fluid or electron-positron plasmas is relatively straightforward. As in R-MHD, the continuity equations are unchanged while we introduce regularization terms in the velocity equations for each species ($l = i, e$ with charges $q_l$ and masses $m_l$). In addition to the vortical twirl term $\bfw_l \times (\grad \times \bfw_l)$ analogous to the one in R-MHD, we add a magnetic twirl term $(q_l/m_l)\bfB \times (\grad \times \bfw_l)$ with a {\it common} coupling strength $\la_l^2$. This is similar to the universal coupling of charged particles to both electric and magnetic fields through the electric charge. Here $\la_l$ are (possibly different) regularizing lengths for the two species. The two twirl terms are obtained by a judicious replacement of $\bfw_l$ by $\bfw_l + q_l \bfB/m_l$ in R-MHD. The combination $\bfw + q \bfB/m$ also appears elsewhere, notably in the study of plasmas in non-inertial frames\cite{thyagaraja-mc-clements-2}. The number densities $n_l$ and $\la_l$ must satisfy the constitutive relations $\la_l^2 n_l = C_l$ where $C_l$ must be constant for a conserved energy to exist. These relations are automatic if $\la_{i,e}$ are chosen to be the Debye lengths or skin depths for ions and electrons, where the ideal equations are known to breakdown. Gauss ($\eps_0 \grad \cdot \bfE = \varrho$), Faraday ($\pdr\bfB/\pdr t = - \grad \times \bfE$) and Amp\`ere ($\mu_0 \eps_0 (\pdr \bfE /\pdr t) = \grad \times \bfB - \mu_0 \bfj_*$) laws take their usual forms with charge density given by $\varrho = \sum_l q_l n_l$. However, the `swirl' current $\bfj_* = \bfj_{\rm flow} + \bfj_{\rm twirl}$ differs from the flow current $\bfj_{\rm flow} = \sum_l q_l n_l \bfv_l$  by an additional regularization term $\bfj_{\rm twirl} = \sum_l q_l n_l \la_l^2 \grad \times \bfw_l$. The constitutive relations ensure that $\bfj_{\rm twirl} = \sum_l \grad \times (\grad \times \la_l^2 \bfj_{{\rm flow},l})$ is solenoidal, thus guaranteeing charge conservation: $\pdr_t \varrho + \grad \cdot \bfj_* = 0$. The constitutive relations and modification of current $\bfj_{\rm flow} \mapsto \bfj_*$ are crucial for obtaining a conserved `swirl' energy including a vortical contribution for {\it compressible} barotropic flow:
	\beqs 
	E^* &=& \int \left[ \sum_{l=i,e} \left( \half n_l m_l ({\bfv_l}^2  + \la_l^2 \bfw_l^2) + U_l(n_l m_l) \right) \right.   \cr
	&& \left. +  \frac{\bfB^2}{2\mu_0} + \frac{\eps_0  \bfE^2}{2} \right] d\bfr 
	\quad \text{where} \quad \grad U_l' = \frac{\grad p_l}{m_l n_l}.
	\label{e:swirl-energy}
	\eeqs 
Here $p_l$ are the partial pressures. The positive definiteness of $E^*$ along with the constitutive relations ensure that the kinetic and compressional energies as well as the enstrophy of each species is bounded, thus helping to regularize vortical singularities. We also derive local conservation laws for swirl energy, linear and angular momenta in our regularized two-fluid model. Unlike in the single-fluid case, we do not have analogues of conserved magnetic and cross helicities. When the number densities $n_{i,e}$ and $\la_i = \la_e = \la$  are {\it constants} and the compressional and electric energies are omitted, the above equations reduce to a conservative regularization of {\it incompressible} quasi-neutral two-fluid plasmas. Interestingly, in the incompressible case {\it alone}, if the current in Amp\`ere's law is taken to be $\bfj_{\rm flow}$, we obtain a {\it different} conserved energy that includes terms with both velocity and magnetic field curls: 
	\beqs
	E^*_{\rm inc} &=& \int \left[\sum_{l} \left(\half n m_l \left({\bfv_l}^2  + \la^2 (\grad \times \bfv_l)^2 \right)\right) + \frac{\bfB^2}{2\mu_0} \right. \cr 
	&& \left. + \frac{\la^2}{2 \mu_0} (\grad \times \bfB)^2  \right] d\bfr.
	\eeqs

In Section \ref{s:heirarchy-of-models} a hierarchy of regularized plasma models is considered. In many physically interesting situations [eg. tokamak or many astrophysical plasmas\cite{Wesson,Kulsrud}] it is reasonable to sacrifice the generality of the full two-fluid model and assume quasi-neutrality $(n_i \approx n_e)$ on scales larger than the Debye length $\la_D$ and frequencies less than the plasma frequency $\om_p$. Additionally, in systems such as accretion disks and planetary magnetospheres \cite{Michel}, one may even ignore electron inertia effects (Hall MHD). The passage from our full regularized two fluid model to the corresponding quasi-neutral, Hall and 1-fluid MHD models is achieved via the successive limits $\eps_0 \to 0$ (non-relativistic limit where the displacement current may be ignored), $m_e \to 0$ ($m_e/m_i \ll 1$) and finally electric charge $e \to \infty$ with $\la_e/\la_i \to 1$ ($L \gg \la_D$ and $\om \ll \om_p$). In each case we have a conserved swirl energy guaranteeing boundedness of enstrophy. In the quasi-neutral limit where $c \to \infty$, $\bfE$ is non-dynamical. It is determined from the electron velocity equation rather than from Gauss' law: 
	\beq
	\bfE = - \bfv_{*e} \times \bfB - \frac{\grad p_e}{en} - \frac{m_e}{e} \left( \pdr_t \bfv_e + \bfw_e \times \bfv_{*e} + \half \grad \bfv_e^2 \right)
	\eeq
where $\bfv_{*e} = \bfv_e + \la_e^2 \grad \times \bfw_e$ is the electron swirl velocity. The situation is analogous to the determination of pressure from the divergence of the Euler equation upon passing to incompressible flow by taking the sound speed $c_s \to \infty$. In the regularized Hall model where electron inertia terms are ignored, magnetic helicity $\int \bfA \cdot \bfB \:d\bfr$ is conserved and in the barotropic case, $\bfB$ is frozen into $\bfv_{*e}$. Finally, when $e \to \infty$ ($L \gg \la_D$) we recover the one-fluid R-MHD model ($\bfv \approx \bfv_i \approx \bfv_e$ and $\la_i =\la_e = \la$) with the magnetic field frozen into the swirl velocity $\bfv_*$.

In Section \ref{s:PB-2-fluid} the Poisson bracket (PB) formalism for regularized compressible two-fluid models is discussed. Interestingly our two-fluid equations follow from the PBs introduced by Spencer-Kaufman\cite{spencer-kaufman} and Holm-Kuperschmidt\cite{holm-kuperschmidt} with the swirl energy $E^*$ taken as the Hamiltonian. Whilst R-MHD admits a Hamiltonian formulation with the Landau-Morrison-Greene PBs \cite{landau,morrison-greene}, we have not identified PBs for the quasi-neutral 2-fluid or Hall MHD models. Moreover, unlike the Hamiltonian and equations of motion (EOM), the 2-fluid PBs do not all reduce to the 1-fluid PBs under the above limiting processes.

In Section \ref{s:reg-field-curl-PBs-Hamiltonian} we exploit the above PB formulation to propose a way of regularizing magnetic field gradients in compressible one- and two-fluid plasma models. In standard tearing mode theory \cite{Hazeltine-Meiss,Wesson,Debasis} the magnetic field can have tangential discontinuities associated with current sheets and reconnection. These current density singularities are usually resolved by resistivity; we propose a {\it conservative regularization.} By analogy with the vortical energy densities $(1/2) \la_l^2 \rho_l (\grad \times \bfv_l)^2$ which regularizes velocities we add $(\la_B^2/2\mu_0)(\grad \times \bfB)^2$ to the swirl energy $E^*$ of (\ref{e:swirl-energy}), to prevent $\bfB$ from developing a large curl. Here $\la_B$ is a {\it constant} cut-off length. The equations of motion obtained from this Hamiltonian using the 2-fluid PBs can be put in the same form as before by replacing $\mu_0 \bfj_*$ in Amp\`ere's law with $\mu_0 \bfj_* - \la_B^2 \:\grad \times (\grad \times (\grad \times \bfB))
$. On the other hand, the introduction of such a magnetic curl energy in the 1-fluid Hamiltonian adds $-(\la_B^2/\rho \mu_0) \bfB \times (\grad \times (\grad \times (\grad \times \bfB)))$ on the RHS of the velocity equation upon use of the 1-fluid PBs. In other words, we have a modified Lorentz force term $\bfj_{**} \times \bfB$ where $\mu_0 \bfj_{**} = \grad \times \bfB + \la_B^2 (\grad \times (\grad \times (\grad \times \bfB)))$.
These third derivatives of $\bfB$ could smooth large gradients in current and field across current sheets just as the $u_{xxx}$ term in KdV does across a shock\cite{whitham}. Interestingly, XMHD\cite{kimura-morrison,abdelhamid-kawazura-yoshida} provides an alternate way of regularizing magnetic though not vortical singularities within a 1-fluid setup. Indeed, the XMHD Hamiltonian includes $(\grad \times \bfB)^2$ but not $(\grad \times \bfv)^2$. Moreover, the resulting regularization terms in the velocity and Faraday equations are quite different from ours due to the use of different PBs (see \S \ref{s:reg-field-curl-PBs-Hamiltonian}). Another essential difference is that the XMHD cut-off lengths $d_{i,e}$ (normalized collisionless skin-depths) are assumed constant unlike our local cut-offs $\la_{i,e}$.


Section \ref{s:discussion} presents a discussion of the results obtained. In Appendix \ref{s:minimality} we establish an interesting uniqueness property of our twirl regularization. We do this for compressible barotropic neutral flows and indicate the extension to two-fluid plasmas. More precisely, we show that the twirl term $\la^2 \bfw \times (\grad \times \bfw)$ is unique among local regularization terms that are at most quadratic in $\bfv$ and with at most three spatial derivatives which preserve Galilean, parity and time-reversal symmetries while also admitting a Hamiltonian-Poisson bracket formulation with the standard continuity equation and Landau-Morrison-Greene PBs. The identification and elimination of possible regularization terms is greatly facilitated by working at the level of the Hamiltonian rather than the equations of motion. It allows us to arrive at the vortical energy term $\int (1/2) \la^2 \rho \: \bfw^2 d\bfr$ with $\la^2 \rho$ constant $(> 0)$, as the only positive definite regularization term satisfying the foregoing criteria subject to decaying or periodic boundary conditions.

\section{Regularized compressible 2-fluid plasma equations}
\label{s:reg-eqns-2-fluid-compress}

The dynamical variables of a 2-fluid plasma are: $\bfE$, $\bfB$, ion and electron velocities $\bfv_{i,e}$, number densities $n_{i,e}$ and partial pressures $p_{i,e}$. The number densities satisfy the continuity equations:
	\beq
	\pdr_t n_{l} + \grad \cdot (n_{l} \bfv_{l}) = 0 \quad \text{where} \quad l = i \;\; \text{or} \;\; e.
	\label{e:reg-cont-eqn-2-fluid}
	\eeq
If $q_{i,e}$ denote the ion and electron charges, then the regularized velocity equations are:
	\beqs
	\pdr_t \bfv_l && +  \bfv_l \cdot \grad \bfv_l = - \ov{n_l m_l} \grad p_l + \frac{q_l}{m_l} (\bfE + \bfv_l \times \bfB) \cr
	&& - \la_l^2 \bfw_l \times (\grad \times \bfw_l) - \frac{\la_l^2 q_l}{m_l} \bfB \times (\grad \times \bfw_l).
	\eeqs
The mass densities and vorticities are $\rho_l = m_l n_l$ and $\bfw_l = \grad \times \bfv_l$ while $\la_{i,e}$ are the short distance cut-offs. For barotropic flow, $(\grad p_l)/\rho_l  = \grad h_l$ where $h_l (\rho_l)$ are the specific enthalpies. In this case, the velocity equations may be written as,
	\beqs
	\pdr_t \bfv_l + \bfw_l \times \bfv_l &=& -\grad \sig_l + \frac{q_l}{m_l}  (\bfE + \bfv_l \times \bfB)\cr 
	&& - \la_l^2 \left[ \bfT^w_l + \frac{q_l}{m_l} \bfT^B_l \right]. 
	\label{e:reg-mom-eq-barotropic-Tw-TB}
	\eeqs
Here $\sigma_l = h_l + \half \bfv_l^2$ are the specific stagnation enthalpies. The vortical and magnetic `twirl' regularization terms for each species are denoted $\bfT^w_l = \bfw_l \times (\grad \times \bfw_l)$ and $\bfT^B_l = \bfB \times (\grad \times \bfw_l)$.  As we will see in \S \ref{s:energy-cons-2-fluid}, conservation of energy requires that the strengths $\la_l^2$ of the vortical $\bfT^w_l$ and magnetic $(q_l/m_l) \bfT^B_l$ twirl forces must be the same for a given species. This resembles the universality of the electric charge $q_l$ through which a particle couples to both electric and magnetic fields. The short-distance regulators $\la_{i,e}$ are assumed to satisfy the constitutive relations $\la_l^2 n_l = C_l$ where $C_l$ are constants. We will see that these constitutive relations help to ensure that the EOM admit a conserved energy. Here $\la_{i,e}$ need not be equal (they could, for example, be the ion and electron collisionless skin depths). Yet another way to express the velocity equations is by introducing the swirl velocities $\bfv_{*l} = \bfv_{l} + \la_{l}^2 \grad \times \bfw_{l}$ which allow us to absorb the regularization terms into the vorticity and magnetic Lorentz force terms,
	\beq
	\pdr_t \bfv_l = - \grad \sig_l + \frac{q_l}{m_l} \bfE  + \bfv_{*l} \times \left(\bfw_l + \frac{q_l}{m_l} \bfB \right).
	\label{e:reg-elec-ion-mom-eqn}
	\eeq
We will see that $\bfw_l$ and $\bfB$ often appear in the combination $\bfw_l + q_l \bfB/m_l$ (see, \cite{Ferraro} and also \cite{thyagaraja-mc-clements-2}). In the latter work, it is shown how the vorticity and magnetic fields are intimately linked in non-inertial frames co-moving with a fluid. The evolution equations for vorticities are 	\beqs
	\pdr_t \bfw_l &+& \grad \times (\bfw_l \times \bfv_l) = \frac{q_l}{m_l} \grad \times (\bfE + \bfv_l \times \bfB) \cr 
	&& - \grad \times \left[\la_l^2 \left(\bfT^w_l + \frac{q_l}{m_l} \bfT^B_l \right) \right]
	\label{e:vorticity-eqn-2-fluid}
	\eeqs
while the Faraday and Amp\`ere evolution equations are
	\beqs
	\dd{\bfB}{t} &=& - \grad \times \bfE \quad \text{and} \quad
	\mu_0 \eps_0 \dd{\bfE}{t} = \grad \times \bfB - \mu_0 \bfj_*
	\label{e:maxwell-evol-eqns-2-fluid}
	\eeqs
with $c = 1/{\sqrt{\mu_0 \eps_0}}$. Here the total `swirl' current density $\bfj_*$ is related to the velocities and densities of the two species via the constitutive law
	\beq
	\bfj_* = \bfj_{*i} + \bfj_{*e} \quad \text{where} \quad \bfj_{*i,e} = q_{i,e} n_{i,e} \bfv_{* i,e}.
	\eeq
The regularized ion and electron swirl currents are a sum of flow and twirl currents for each species
	\beq
	\bfj_{*l} = \bfj_{{\rm flow},l} + \bfj_{{\rm twirl},l} \equiv q_l n_l \bfv_l + q_l n_l \la_l^2 \grad \times \bfw_l.
	\label{e:swirl-current-2-fluid}
	\eeq 
The constitutive laws $\la_l^2 n_l = C_l$ allow us to write the twirl currents in manifestly solenoidal form:
	\beq
	\bfj_{{\rm twirl},l} = \grad \times (\grad \times \la_l^2 \bfj_{{\rm flow},l}).
	\label{e:j-twirl-as-curlcurl}
	\eeq
Postulating that the current appearing in Amp\`ere's law is $\bfj_*$ rather than the unregularized $\bfj_{\rm flow}$ allows us to derive a conserved energy (\ref{e:energy-density-2-fluid}) in \S~\ref{s:energy-cons-2-fluid}. In addition, the electric and magnetic fields must satisfy
	\beq
	\grad \cdot \bfB = 0 \;\; \text{and} \;\; \eps_0 \grad \cdot \bfE = \varrho \;\; \text{where} \;\;\varrho =  n_i q_i + n_e q_e
	\eeq
is the charge density. The consistency of the inhomogeneous Maxwell equations require that $\bfj_*$ and $\varrho$ satisfy the local conservation law $\pdr_t \varrho + \grad \cdot \bfj_* = 0$. Our regularized current does indeed satisfy this condition since $\grad \cdot \bfj_{\rm twirl} = 0$ and by the continuity equations,
	\beq
	\grad \cdot \bfj_{\rm flow} = \grad \cdot \sum_l q_l n_l \bfv_l = - \pdr_t \sum_l q_l n_l = - \pdr_t \varrho.
	\eeq

\subsection{Local conservation laws}

In this section, we show that the compressible regularized 2-fluid equations of \S \ref{s:reg-eqns-2-fluid-compress} possess locally conserved energy,  linear and angular momenta and identify the corresponding currents. The conservation of energy depends crucially on the constitutive relations and the modification of Amp\`ere's law to include a regularized `twirl' current in addition to the flow current (\ref{e:swirl-current-2-fluid}). In the limit of constant densities $n_{i,e}$ we obtain a locally conserved energy for incompressible 2-fluid plasmas provided the regularization lengths $\la_{i,e}$ are equal. Interestingly, we discover {\it another} way of regularizing the incompressible equations, the difference being that it is $\bfj_{\rm flow}$ and  not $\bfj_*$ that appears in Amp\`ere's law. The resulting conserved energy shows that velocity as well as field curls are regularized. However, this approach does not generalize to the compressible case. Unlike in ideal and twirl regularized 1-fluid MHD, magnetic helicity $\int \bfA \cdot \bfB \; d\bfr$ is {\it not} conserved in the general 2-fluid model. However, it {\it is} conserved in the Hall 2-fluid limit where electron inertia terms are ignored (\S \ref{s:R-Hall-2fluid}). On the other hand, we do not have a 2-fluid analogue of the conserved cross helicity of the (regularized) 1-fluid MHD equations.

\subsubsection{Local conservation of energy}
\label{s:energy-cons-2-fluid}

The regularized equations (\ref{e:reg-cont-eqn-2-fluid}), (\ref{e:reg-mom-eq-barotropic-Tw-TB}) and (\ref{e:maxwell-evol-eqns-2-fluid}) for barotropic 2-fluid plasmas obeying the constitutive laws $\la_l^2 n_l = C_l$ possess a positive definite swirl energy density
	\beq
	{\cal E}^* = \sum_{l=i,e} \left[\half \rho_l ({\bfv_l}^2  + \la_l^2 \bfw_l^2) + U(\rho_l)\right] + \frac{\bfB^2}{2\mu_0} + \frac{\eps_0}{2} \bfE^2
	\label{e:energy-density-2-fluid}
	\eeq
satisfying a local conservation law $\pdr_t {\cal E}^* + \grad \cdot \bff = 0$ where
	\beqs
	\bff &=&  \sum_{l} \left[\sig_l \rho_l \bfv_l 
	   + \la_l^2 \rho_l \bfw_l \times \left( \bfv_l \times \bfw_l + \frac{q_l}{m_l} (\bfE + \bfv_l \times \bfB ) \right. 
	\right.  \cr
	&&  \left. \left. - \la_l^2 \left(\bfT^w_l + \frac{q_l}{m_l} \bfT^B_l \right) \right) \right] + \frac{\bfE \times \bfB}{\mu_0}.
	\label{e:energy-cons-2-fluid}
	\eeqs
With appropriate BCs (E.g. decaying or periodic) the total swirl energy $\int {\cal E}^* d\bfr$ is a constant of motion. Thus in addition to the kinetic and potential energies of each species, their enstrophies $\int \bfw_l^2 d\bfr$ (or vortical energies) are bounded above. The corresponding kinetic, vortical and potential energy densities in ${\cal E}^*$ will be denoted ${\cal KE}, {\cal VE}$ and ${\cal PE}$. The energy flux may be compactly written in terms of the swirl velocities $\bfv_{l*}$:
	\beqs
	\bff &=& \sum_{l} \left[\sig_l \rho_l \bfv_l + \bfE \times \left(\frac{\bfB}{\mu_0} - \grad \times \la_l^2 \bfj_{{\rm flow},l} \right) \right. \cr
	&& \left. + \la_l^2 \rho_l \bfw_l \times \left( \bfv_{l*} \times \left(\bfw_l + \frac{q_l}{m_l} \bfB \right) \right) \right].
	\label{e:energy-current-2flu-compress}
	\eeqs
The first term comes from ideal flow while the second is the Poynting flux, which is augmented by a regularizing term. It may be noted that the combination $\bfB - \mu_0 \grad \times \la_l^2 \bfj_{{\rm flow},l}$ also appears in Amp\`ere's law (\ref{e:maxwell-evol-eqns-2-fluid}).

Let us sketch the proof of (\ref{e:energy-cons-2-fluid}), which involves some remarkable cancellations. To begin we take the dot product of the velocity equations (\ref{e:reg-elec-ion-mom-eqn}) for each species with $\rho_l \bfv_l$. Since the vorticity and magnetic forces do no work,
	\beqs
	\half \rho_l \pdr_t \bfv_l^2 &=& -\rho_l \bfv_l \cdot \grad \left( h_l + \half \bfv_l^2 \right) + n_l q_l \bfv_l \cdot \bfE \cr
	&& - \la_l^2 \rho_l \bfv_l \cdot \bfT^w_l - \la_l^2 n_l q_l \bfv_l \cdot \bfT^B_l
	\eeqs
for each $l = i, e$.
Using (\ref{e:reg-cont-eqn-2-fluid}) we get
	\beqs
	\pdr_t({\cal KE}_l) &+& \half \bfv_l^2 \grad \cdot (\rho_l \bfv_l) + \rho_l \bfv_l \cdot \grad \left( h_l + \half \bfv_l^2 \right)  \cr
	&=& n_l q_l \bfv_l \cdot \bfE - \la_l^2 \rho_l \bfv_l \cdot \left[ \bfT^w_l + \frac{q_l}{m_l} \bfT^B_l \right].
	\eeqs
Again by the continuity equation, 
	\beqs
	\rho_l \bfv_l \cdot \grad h_l &=& \grad \cdot (\rho_l h_l \bfv_l) - U_l'(\rho_l) \grad \cdot (\rho_l \bfv_l) \cr
	&=& \grad \cdot (\rho_l h_l \bfv_l) + \pdr_t U_l.
	\eeqs
Thus the time derivative of the sum of kinetic and potential energy densities of each species is
	\beqs
	\pdr_t({\cal KE}_l + {\cal PE}_l) &=& - \grad \cdot (\sig_l  \rho_l \bfv_l ) + n_l q_l \bfv_l \cdot \bfE \cr
	&& - \la_l^2 \rho_l \bfv_l \cdot \left( \bfT^w_l + \frac{q_l}{m_l} \bfT^B_l \right).
	\label{e:time-der-of-KE_l+PE_l-2-fluid}
	\eeqs
The second term on the RHS is the work done by $\bfE$. To write the work done by the twirl regularization forces in conservation form and introduce the  vortical energy density, we dot the vorticity evolution equation (\ref{e:vorticity-eqn-2-fluid}) for each species with $\la_l^2 \rho_l \bfw_l$:
	\beqs
	\pdr_t \left( {\cal VE}_l \right) 
	&=& \la_l^2 \rho_l \bfw_l \cdot \grad \times \left[ (\bfv_l \times \bfw_l)  
	+ \frac{q_l}{m_l} (\bfE + \bfv_l \times \bfB) \right. \cr
	&& \left. -  \la_l^2  \left(\bfT^w_l + \frac{q_l}{m_l} \bfT^B_l \right) \right].
	\label{e:enstrophic-energy-time-der-2-fluid}
	\eeqs
The vector identity for the divergence of a cross product allows us to write (\ref{e:enstrophic-energy-time-der-2-fluid}) as
	\beqs
	\pdr_t({\cal VE}_l) 
	&=& \la_l^2 \rho_l \left[ (\bfv_l \times \bfw_l) 
	+ \frac{q_l}{m_l} (\bfE + \bfv_l \times \bfB )\right.\cr
	&& \left. - \la_l^2 \left(\bfT^w_l + \frac{q_l}{m_l} \bfT^B_l \right) \right] \cdot \grad \times \bfw_l \cr
		&& + \la_l^2 \rho_l \grad \cdot \left[ \left( \bfv_l \times \bfw_l + \frac{q_l}{m_l} (\bfE + \bfv_l \times \bfB \right. \right. \cr
	&& \left. \left. -  \la_l^2 \left(\bfT^w_l + \frac{q_l}{m_l} \bfT^B_l \right) \right)\times \bfw_l \right].
	\eeqs
Using the properties of the scalar triple product and rearranging, the rate of change of vortical energy density of each species is
	\beqs
	&&( {\cal VE}_l )_t 
	= \la_l^2 \rho_l \bfv_l \cdot \left[ \left(\bfw_l + \frac{q_l}{m_l} \bfB \right) \times \grad \times \bfw_l \right] \cr
	&& + \la_l^2 \rho_l \grad \cdot \left[ \left[ \bfv_l \times \bfw_l + \frac{q_l}{m_l} (\bfE + \bfv_l \times \bfB) \right. \right. \cr
	&& \left. \left. -  \la_l^2 \left[\bfT^w_l + \frac{q_l}{m_l} \bfT^B_l \right] \right] \times \bfw_l \right] + \bfE \cdot \grad \times (\la_l^2 n_l q_l \bfw_l) .
	\label{e:time-der-of-EE_l-2-fluid}
	\eeqs
We add (\ref{e:time-der-of-KE_l+PE_l-2-fluid}) and (\ref{e:time-der-of-EE_l-2-fluid}), sum over species and identify the swirl current $\bfj_*$ from (\ref{e:swirl-current-2-fluid}). The work done by the twirl forces $\la_l^2 \rho_l \bfv_l \cdot (\bfT^w_l + (q_l/m_l) \bfT^B_l)$ cancels out giving:
	\beqs
	&&\pdr_t ({\cal KE} + {\cal PE} + {\cal VE}) 
	+ \sum_l \grad \cdot \left[ \sig_l \rho_l \bfv_l - \la_l^2  \rho_l \left(  \bfv_l \times \bfw_l  \right. \right. \cr
	 && \left. \left.  - \frac{q_l}{m_l} (\bfE + \bfv_l \times \bfB ) + \la_l^2 \left(\bfT^w_l + \frac{q_l}{m_l} \bfT^B_l \right) \right) \times \bfw_l \right] \cr
	&& = \bfE \cdot \bfj_*.
	\eeqs
Now we use the regularized Maxwell equations (\ref{e:maxwell-evol-eqns-2-fluid}) to calculate the total work done by the electric field
	\beqs
	\bfE \cdot \bfj_* &=& \frac{\bfE  \cdot (\grad \times \bfB)}{\mu_0} - \eps_0 \bfE \cdot \pdr_t \bfE = \frac{\bfB \cdot \grad \times \bfE}{\mu_0} \cr
	&& + \grad \cdot \left( \frac{\bfB \times \bfE}{\mu_0} \right) - \pdr_t \left( \frac{\eps_0 \bfE^2}{2} \right) \cr 
	&=& -\pdr_t \left( \frac{\eps_0 \bfE^2}{2} + \frac{\bfB^2}{2 \mu_0} \right) + \grad \cdot \left( \frac{\bfB \times \bfE}{\mu_0} \right).
	\eeqs
Evidently it is crucial that the current in Amp\`ere's law is taken as the swirl current $\bfj_*$ instead of $\bfj_{\rm flow}$ to obtain the local conservation law for swirl energy ${\cal E}^*$ (\ref{e:energy-density-2-fluid}).

\subsubsection{Conservation of energy in incompressible flow and regularization of $\bfB$}
\label{s:energy-cons-2-fluid-incompress}

For low acoustic Mach numbers $(M_l = |\bfv_l/c^s_{l}|) \ll 1$, the number densities $n_l$ are spatially and temporally constant to leading order. In this limit, the plasma motions while producing changes in $\bfE$ and $\bfB$ do not produce propagating EM waves. This is equivalent to dropping the displacement current in Maxwell's equations ($c \gg c^s_l$). For physical consistency we must take $\eps_0 \to 0$.

By taking $n_{i,e}$ and the regularizing lengths $\la_{i,e}$ to be constants and $\eps_0 \to 0$ we arrive at an incompressible 2-fluid model. The continuity equations become $\grad \cdot  \bfv_{i,e} = 0$ and $\eps_0 \to 0$ in Gauss' law implies quasi-neutrality $(n_i \approx n_e \equiv n$, assuming $q_i = - q_e)$. The velocity equations are 		
	\beqs
	\pdr_t \bfv_l + \bfw_l \times \bfv_l &=& - \grad \sigma_l + \frac{q_l}{m_l} (\bfE + \bfv_l \times \bfB) \cr
	&& - \la_l^2 \left(\bfw_l + \frac{q_l}{m_l} \bfB \right) \times (\grad \times \bfw_l)
	\label{e:reg-mom-eqn-incompress-2flu}
	\eeqs
where $\sigma_l = p_l/\rho_l + \half \bfv_l^2$ for $l = i, e$. In this limit Amp\`ere's law (\ref{e:maxwell-evol-eqns-2-fluid}) becomes $\grad \times \bfB = \mu_0 \bfj_*$. It follows from \S \ref{s:energy-cons-2-fluid} that upon dropping compressional and electric energies, the energy density, 
	\beq
	{\cal E}^*_{\rm inc} = \sum_{l} \left[\half \rho_l ({\bfv_l}^2  + \la_l^2 \bfw_l^2) \right] + \ov{2 \mu_0} \bfB^2
	\label{e:energy-density-2-fluid-incompressible}
	\eeq
satisfies a local conservation law with the energy current of (\ref{e:energy-current-2flu-compress}). As a consequence, the enstrophy of each species is bounded and velocity curls cannot become too large though there is no \`a priori bound on field curls.

Remarkably, (as indicated in Ref.\cite{thyagaraja-2}) there is another way  of defining the regularized {\it incompressible} 2-fluid model (with $\la_i = \la_e = \la$) where the field gradient $\grad \times \bfB$ is also regularized along with $\grad \times \bfv$. This is achieved by keeping the velocity (\ref{e:reg-mom-eqn-incompress-2flu}) and Faraday equations unchanged but postulating that the current in Amp\`ere's law is the flow current $\bfj_{\rm flow} = n \sum_l q_l \bfv_l$ rather than the swirl current $\bfj_*$ (\ref{e:swirl-current-2-fluid}),
	\beq
	\grad \times \bfB = \mu_0 \, \bfj_{\rm flow} 
	\label{e:Ampere-Faraday-2fluid-jflow}.
	\eeq
Under these circumstances, we find a new conserved energy density 
	\beqs
	\tl {\cal E}^*_{\rm inc} = \sum_{l}\left[ \frac{\rho_l}{2} ({\bfv_l}^2  + \la^2 \bfw_l^2) \right] + \frac{\bfB^2}{2\mu_0} 
	+ \frac{\la^2(\grad \times \bfB)^2}{2 \mu_0} \quad
	\label{e:energy-density-2-fluid-incompress-jflow}
	\eeqs
and associated flux 
	\beqs
	&& \tl \bff = \sum_{l} \left[\sig_l \rho_l \bfv_l + \la^2 \rho_l \bfw_l \times \left( \bfv_{l*} \times \left(\bfw_l + \frac{q_l}{m_l} \bfB \right) \right)\right] \cr
	&&  + \frac{\bfE \times \bfB}{\mu_0} + \la^2 \left[ \bfE \times (\grad \times \bfj_{\rm flow}) - \bfj_{\rm flow} \times (\grad \times \bfE) \right]
	\label{e:energy-current-incomp-curlB-curlE}
	\eeqs
satisfying a local conservation law $\pdr_t \tl {\cal E}^*_{\rm inc} + \grad \cdot \tl \bff = 0$. This regularization of incompressible flow is remarkable in that the $L^2$ norms of $\bfv, \bfB, \grad \times \bfv$ and $\grad \times \bfB$  are all bounded (say with decaying/periodic BCs). Since in addition, $\grad \cdot \bfv_{i,e} = \grad \cdot \bfB = 0$, we expect vortical singularities as well as singularities in magnetic field gradients to be regularized in this model. The $L^2$-norm of $\bfj_{\rm flow}$ is also bounded as a consequence of Amp\`ere's law (\ref{e:Ampere-Faraday-2fluid-jflow}).

To derive Eqs.~(\ref{e:energy-density-2-fluid-incompress-jflow}) and (\ref{e:energy-current-incomp-curlB-curlE}) we dot the velocity equations (\ref{e:reg-mom-eqn-incompress-2flu}) for each species with $\rho_l \bfv_l$ to get,
	\beq
	\frac{\rho_l}{2} \pdr_t \bfv_l^2 = -\rho_l \bfv_l \cdot \grad \sig_l + n q_l \bfv_l \cdot \bfE - \la_l^2 \rho_l \bfv_l \cdot \left[ \bfT^w_l + \frac{q_l}{m_l} \bfT^B_l \right].
	\eeq
As $\rho_l$ are constants and $\grad \cdot \bfv_l = 0$,
	\beq
	( {\cal KE}_l )_t + \grad \cdot (\sig_l  \rho_l \bfv_l ) = \bfj_{{\rm flow},l} \cdot \bfE - \la_l^2 \rho_l \bfv_l \cdot \left[ \bfT^w_l + \frac{q_l}{m_l} \bfT^B_l \right].
	\label{e:time-der-of-KE_l+PE_l-2-fluid-incompress}
	\eeq
To introduce the vortical energy density, we dot the curl of (\ref{e:reg-mom-eqn-incompress-2flu}) for each species with $\la_l^2 \rho_l \bfw_l$ to get
	\beqs
	({\cal VE}_l)_t &=& \la_l^2 \rho_l \bfw_l \cdot \grad \times \left[ (\bfv_l \times \bfw_l)  
	+ \frac{q_l}{m_l} (\bfE + \bfv_l \times \bfB) \right. \cr
	&& \left. -  \la_l^2  \left(\bfT^w_l + \frac{q_l}{m_l} \bfT^B_l \right) \right].
	\eeqs
Vector identities allow us to write
	\beqs
	 ({\cal VE}_l)_t
	&=& \la_l^2 \rho_l \bfv_l \cdot \left\{ \left(\bfw_l + \frac{q_l}{m_l} \bfB \right) \times \grad \times \bfw_l \right\} \cr
	&& + \bfE \cdot \grad \times (\la_l^2 n q_l \bfw_l) \cr
	&& + \la_l^2 \rho_l \grad \cdot \left[ \left( \bfv_l \times \bfw_l + \frac{q_l}{m_l} \left(\bfE + \bfv_l \times \bfB  \right) \right.  \right. \cr
	&& \left. \left.  - \la_l^2  \left(\bfT^w_l + \frac{q_l}{m_l} \bfT^B_l \right) \right) \times \bfw_l \right].
	\label{e:time-der-of-EE_l-2-fluid-incompress}
	\eeqs
Adding (\ref{e:time-der-of-KE_l+PE_l-2-fluid-incompress}) and (\ref{e:time-der-of-EE_l-2-fluid-incompress}) and summing over species we get 
	\beqs
	&& \pdr_t ({\cal KE} + {\cal VE}) 
	 + \grad \cdot  \sum_l \left[ \sig_l \rho_l \bfv_l + \la_l^2 \rho_l \bfw_l \times \left( \bfv_l \times \bfw_l \right. \right.\cr
	&& \left. \left. + \frac{q_l}{m_l} (\bfE + \bfv_l \times \bfB ) 
	- \la_l^2 \left(\bfT^w_l + \frac{q_l}{m_l} \bfT^B_l \right) \right) \right] \cr
	&& \qquad \qquad = \: \bfE \cdot \left[ \bfj_{\rm flow} + \bfj_{\rm twirl} \right].
	\label{e:incompr-dt-of-ke+ee}
	\eeqs
where $\bfj_{\rm twirl} = \sum_l \grad \times \grad \times \la_l^2 \bfj_{{\rm flow},l}$ (\ref{e:j-twirl-as-curlcurl}). The work done by $\bfE$ is got from (\ref{e:Ampere-Faraday-2fluid-jflow}) (abbreviating flow and twirl):
	\beqs
	\bfE \cdot \bfj_{\rm fl} 
	&=& -\pdr_t \left( \frac{\bfB^2}{2 \mu_0} \right) + \grad \cdot \left( \frac{\bfB \times \bfE}{\mu_0} \right) \quad \text{and} \cr
	\bfE \cdot \bfj_{\rm tw} &=&
	\sum_l \left[\grad \times \la_l^2 \bfj_{{\rm fl}, l} \cdot \grad \times \bfE  - \grad \cdot ( \bfE \times \grad \times \la_l^2\bfj_{{\rm fl}, l} ) \right] \cr
	&=& \sum_l \left[\la_l^2 \, \bfj_{{\rm fl},l} \cdot \grad \times (\grad \times \bfE)  \right. \cr
	&+& \left. \grad \cdot \left(\la_l^2 \bfj_{{\rm fl},l} \times (\grad \times \bfE) - \bfE \times \grad \times \la_l^2\bfj_{{\rm fl}, l} \right) \right]. \qquad
	\label{e:electric-fld-dot-j-twirl}
	\eeqs
If we assume $\la_i = \la_e = \la$ (constant) then $\sum_l \la_l^2 \bfj_{{\rm fl},l} = \la^2 \bfj_{\rm fl} = (\la^2/\mu_0) \grad \times \bfB$, so that $\bfE \cdot \bfj_{\rm twirl}$ becomes 
	\beq
	- \left( \frac{\la^2 (\grad \times \bfB)^2 }{2 \mu_0} \right)_t + \grad \cdot (\la^2 \bfj_{\rm fl} \times (\grad \times \bfE) - \bfE \times \grad \times \la^2 \bfj_{\rm fl}).
	\label{e:time-der-constant-la-2-fluid-incompress}
	\eeq
Putting this in (\ref{e:incompr-dt-of-ke+ee}) we get the conservation of energy $\tl {\cal E}^*_{\rm inc}$ (\ref{e:energy-density-2-fluid-incompress-jflow}). Notably this trick of replacing $\bfj_*$ by $\bfj_{\rm flow}$ in Amp\`ere's law does {\it not} lead to a conserved energy for compressible flow: $\la_{i,e}$ are not constants and cannot be taken inside the derivatives in (\ref{e:time-der-constant-la-2-fluid-incompress}) to obtain a conserved energy including $(\grad \times \bfB)^2$. As mentioned in \S \ref{s:energy-cons-2-fluid}, for compressible flow, we must include the twirl current in Amp\`ere's law  to obtain the conserved swirl energy (\ref{e:energy-density-2-fluid}).

\subsubsection{Local conservation of linear and angular momenta}

Returning to the compressible 2-fluid equations, we obtain a local conservation law $\pdr_t {\cal P}^\al + \pdr_\beta \Pi^{\al \beta} = 0$ for the total momentum density $\vec {\cal P} = \vec {\cal P}_{\rm mech} + \vec {\cal P}_{\rm field} = \sum_{l} \rho_l \bfv_l + \eps_0 (\bfE \times \bfB)$ and symmetric stress tensor,
	\beqs
	&& \Pi^{\al \beta} = p \del^{\al \beta} + \sum_{l} \left[\rho_l v_l^\al v_l^\beta + \la_l^2 \rho_l \left( \frac{\bfw_l^2}{2} \del^{\al \beta} - w_l^\al w_l^\beta \right) \right] \cr
	&& + \ov{\mu_0}\left( \frac{\bfB^2}{2} \del^{\al \beta} - B^\al B^\beta \right) + \eps_0 \left( \frac{\bfE^2}{2} \del^{\al \beta} - E^\al E^\beta \right).
	\label{e:mom-flux}
	\eeqs
Here $p = p_i + p_e$. The first and last pairs of terms, $\Pi^{\al \beta}_{\rm Euler}$ and $\Pi^{\al \beta}_{\rm field}$ in the flux are familiar from ideal flow and the Poynting flux of electrodynamics. The vortical regularization term in between is similar to the latter with the constants $\la_l^2 \rho_l$ playing the role of $\ov{{\mu_0}}$ and $\eps_0$.

To obtain (\ref{e:mom-flux}), we first multiply the continuity equation (\ref{e:reg-cont-eqn-2-fluid}) by $m_l \bfv_l$ and velocity equation (\ref{e:reg-elec-ion-mom-eqn}) by $\rho_l = n_{l} m_{l}$, add them and sum over species to get
	\beqs
	&& \sum_l \left[ (\rho_l \bfv_l)_t + \rho_l (\bfv_l \cdot \grad \bfv_l) + m_l \bfv_l  \grad \cdot (n_l \bfv_l) \right]\cr	&=& - \grad p  + \sum_l \left[ n_l q_l (\bfE + \bfv_l \times \bfB) - \la_l^2 \rho_l \bfw_l \times (\grad \times \bfw_l) \right. \cr
	&& \left. - \la_l^2 n_l q_l \bfB \times (\grad \times \bfw_l) \right]. 
	\eeqs
Using Gauss' law, $\eps_0 \grad \cdot \bfE = \sum_l n_l q_l $ and the formulae for flow and twirl currents (\ref{e:swirl-current-2-fluid}) we get
	\beqs
	 && \pdr_t {\cal P}_{\rm mech}^\al + \pdr_\beta \sum_l (\rho_l v_l^\al v_l^\beta) = - \grad^\al p  + \eps_0 E^\al (\grad \cdot \bfE) 
	\cr
	&& + (\bfj_* \times \bfB)^\al - \sum_l \la_l^2 \rho_l (\bfw_l \times (\grad \times \bfw_l))^\al.
	\eeqs
From Amp\`ere's law $\mu_0 \bfj_* \times \bfB = (\grad \times \bfB) \times \bfB -  \mu_0 \eps_0(\pdr_t \bfE) \times \bfB$ and Faraday's law we get
	\beqs
	\pdr_t {\cal P}_{\rm mech}^\al + \pdr_\beta \Pi^{\al \beta}_{\rm Euler} &=& \eps_0 E^\al \: \grad \cdot \bfE -\ov{\mu_0} (\bfB \times (\grad \times \bfB))^\al 
	\cr
	&& - \eps_0 (\pdr_t ( \bfE \times \bfB) + \bfE \times (\grad \times \bfE))^\al 
	\cr
	&& - \sum_l \la_l^2 \rho_l (\bfw_l \times (\grad \times \bfw_l))^\al.
	\eeqs
Using the identity $(\bfS \times (\grad \times \bfS))^\al = \half \pdr^\al \bfS^2 - S^\beta \pdr^\beta S^\al$ and solenoidal nature of $\bfB$ and $\bfw$ we get 
	\beqs
	\pdr_t {\cal P}^\al &+& \pdr_\beta \left[ \Pi^{\al \beta}_{\rm Euler} + \ov{\mu_0}\left(\frac{\bfB^2}{2} \del^{\al \beta} - B^\al B^\beta \right) \right. \cr
	&& \left. + \sum_l  \la_l^2 \rho_l \left(\frac{\bfw_l^2}{2} \del^{\al \beta} - w_l^\al w_l^\beta \right) \right] 
	\cr &&
	= \eps_0 \left[ E^\al (\grad \cdot \bfE) - \half \pdr^\al \bfE^2  + E^\beta \pdr^\beta E^\al \right]
	\eeqs
which implies the local conservation law (\ref{e:mom-flux}).

The time derivative of angular momentum density $\vec {\cal L} =  \bfr \times \vec {\cal P} =  {\bf r} \times  \left( \sum_l \rho_l \bfv_l + \eps_0 \bfE \times \bfB \right)$ is calculated using the local conservation law for momentum density and the symmetry of $\Pi_{\al \beta} (\ref{e:mom-flux})$:
	\beq
	\dd{{\cal L}_\al}{t} = \eps_{\al \beta \gamma} r_\beta \pdr_t {\cal P}^\gamma
	= - \eps_{\al \beta \gamma} r_\beta \pdr_\eta \Pi_{\gamma \eta} = - \pdr_\eta \Lambda_{\al \eta}.
	\eeq
Thus $\pdr {\cal L}_\al/ \pdr t + \pdr_\beta \Lambda_{\al \beta} = 0$ where $\Lambda_{\al \beta} = \eps_{\al \gamma \del} r_\gamma \Pi_{\del \beta} $ is the angular momentum flux tensor.

\section{Hierarchy of regularized models}
\label{s:heirarchy-of-models}

The regularized compressible $2$-fluid plasma equations have several free parameters $\eps_0, m_e/m_i$, electric charge $e$ and $\la_i/\la_e$. By successively taking $\eps_0 \to 0$, $m_e/m_i \to 0$ and $e \to \infty$ together with $\la_i / \la_e \to 1$ we get the (regularized) quasi-neutral 2-fluid, Hall and 1-fluid MHD models.

\subsection{Regularized quasi-neutral 2-fluid plasma}
\label{s:quasi-neutral-2-fluid}

For quasi-neutral plasmas with $q_i = -q_e = e$, the number densities of ions and electrons are approximately equal, $n_i \approx n_e = n$. The equations of such a plasma may be formally obtained from the compressible 2-fluid model (\S \ref{s:reg-eqns-2-fluid-compress}) by taking $\eps_0 \to 0$. Indeed, if $n_i, n_e \to n$, Gauss' law $\grad \cdot \bfE = e (n_i - n_e)/\eps_0$ seems to suggest that $\grad \cdot \bfE = 0$. But in fact, the electric field is not divergence free (especially on length scales comparable to the Debye length). We must also let $\eps_0 \to 0$ in such a way that $e (n_i - n_e)/\eps_0$ has a finite limit. The limit $\eps_0 \to  0$ is a convenient way of taking the non-relativistic limit $c = 1/\sqrt{\eps_0 \mu_0} \to \infty$ ($\mu_0$ is a constant) in which $v_{i,e}/c \ll 1$ in the lab frame. In this limit $\bfE$ is not a propagating degree of freedom and we may ignore the displacement current term in Amp\`ere's law (as stated in \ref{s:energy-cons-2-fluid-incompress}). Furthermore, $\bfE$ is no longer determined by Gauss' law but obtained from the electron velocity equation as discussed below.

In the non-relativistic quasi-neutral limit $\eps_0 \to 0$, the Faraday and Amp\`ere-Maxwell equations become
	\beq
	\grad \cdot \bfB = 0, \quad \dd{\bfB}{t} = - \grad \times \bfE \quad \text{and} \quad
	\grad \times \bfB = \mu_0 \bfj_*.
	\label{e:Maxwell-eqn-quasi-neutral}
	\eeq
For consistency, $\grad \cdot \bfj_*$ must vanish as we will verify using the continuity equations 
	\beq
	\pdr_t n + \grad \cdot (n \bfv_{i,e}) = 0.
	\label{e:cont-eqn-quasineutral}
	\eeq
The difference between the continuity equations gives
	\beq
	\grad \cdot n( \bfv_i - \bfv_e) = 0.
	\eeq
Multiplying by $e$, we see that the flow current $\bfj_{{\rm flow}} = e n (\bfv_i -\bfv_e)$ is solenoidal. On the other hand, the twirl current $\bfj_{\rm twirl} = \sum_l \grad \times (\grad \times \la_l^2 \bfj_{{\rm flow},l})$ is always divergence free, so the total current $\bfj_* = \bfj_{\rm flow} + \bfj_{\rm twirl}$ for quasi-neutral plasmas is solenoidal. This also follows from the Amp\`ere-Maxwell equation when $\eps_0 \to 0$.

The ion and electron velocity equations $(l = i, e)$ for quasi-neutral plasmas are
	\beq
	\pdr_t \bfv_l + \bfw_l \times \bfv_{*l} = - \frac{\grad p_l}{m_l n} - \frac{\grad \bfv_l^2}{2} \pm \frac{e}{m_l} (\bfE + \bfv_{*l} \times \bfB).
	\label{e:reg-ion-elec-mom-quasineutral}
	\eeq
$\bfE$ is determined from the electron velocity equation:
	\beq
	\bfE_{\rm qn} = -  \bfv_{*e} \times \bfB - \frac{\grad p_e}{en} - \frac{m_e}{e}\left [ \pdr_t \bfv_e + \bfw_e \times \bfv_{*e} + \frac{ \grad \bfv_e^2}{2} \right].
	\label{e:E-from-e-mom-eqn}
	\eeq
The relation between general and quasi-neutral 2-fluid plasmas bears a resemblance to that between compressible and incompressible  barotropic neutral flows. In compressible flow, pressure $p$ is obtained from density $\rho$ using the barotropic relation. Similarly, in general 2-fluid plasmas $\bfE$ is determined in terms of the charge density from Gauss' law. On the other hand, in the incompressible ($\grad \cdot \bfv = 0$) constant density $(\rho = \rho_0)$ limit, $p$ is no longer determined by the barotropic relation but from the Poisson equation $ [\grad^2 p = -\rho_0 \grad \cdot (\bfv \cdot \grad \bfv)]$ obtained by taking the divergence of the velocity equation. Similarly, in quasi-neutral plasmas, $\bfE$ is determined from the electron velocity equation rather than from Gauss' law. Moreover, $\eps_0 \to 0$ ($c \to \infty$) is like taking the Mach number to zero (sound speed $c_s \to \infty$).

In this limit, the electric term drops out of the conserved swirl energy for barotropic flow generalizing (\ref{e:energy-density-2-fluid-incompressible}):
	\beq
	{\cal E}_{\rm qn}^* = \sum_{l = i, e} \left(\frac{\rho_l{\bfv_l}^2}{2} + U_l(\rho_l) + \frac{\la_l^2 \rho_l \bfw_l^2}{2} \right) + \frac{\bfB^2}{2\mu_0}.
	\eeq
Here $\rho_l = m_l n$ and $\grad U_{l}' = \grad h_l = \grad p_l/\rho_l$ for $l = i,e$.

\subsection{Regularized Hall MHD without electron inertia}
\label{s:R-Hall-2fluid}

In the limit $m_e/m_i \ll 1$ we drop electron inertia terms to get the regularized Hall model. The Maxwell equations, continuity equations and ion velocity equation are as in the quasi-neutral theory (\S \ref{s:quasi-neutral-2-fluid}). In (\ref{e:E-from-e-mom-eqn}) we drop electron inertia terms to get
	\beq
	\bfE_{\rm Hall} = -  \bfv_{*e} \times \bfB - \frac{\grad p_e}{en}
	\label{e:hall-electric-field}
	\eeq
For barotropic flow, where $\grad p_e/n$ is a gradient, Faraday's law becomes $\pdr_t \bfB = \grad \times (\bfv_{*e} \times \bfB)$. Thus unlike in the full 2-fluid model, in the R-Hall model the magnetic field is frozen into the electron swirl velocity.

We have an additional conserved quantity: magnetic helicity satisfies the local conservation law
	\beq
	\pdr_t(\bfA \cdot \bfB) + \grad \cdot \left( \phi \bfB + \bfE_{\rm Hall} \times \bfA - \frac{2 \tl h_e \bfB}{e} \right) = 0.
	\label{e:mag-hel-loc-cons-Hall-2fluid}
	\eeq
Here $\phi$ is the scalar potential and we assume the barotropic condition $(\grad p_e)/n = \grad \tl h_e$. To obtain (\ref{e:mag-hel-loc-cons-Hall-2fluid}), we use the homogeneous Maxwell equations and $\bfE = -\grad \phi - \pdr_t \bfA$ to compute
	\beqs
	(\bfA \cdot \bfB)_t &=&  - \bfB \cdot \grad \phi - \bfB \cdot \bfE - \bfA \cdot \grad \times \bfE 
	\cr
	&=& - \grad \cdot ( \phi \bfB + \bfE \times \bfA) - 2 \bfE \cdot \bfB.
	\eeqs
Using the quasi-neutral electric field (\ref{e:E-from-e-mom-eqn}) we get
	\beqs
	(\bfA \cdot \bfB)_t &=& - \grad \cdot (\phi \bfB + \bfE_{\rm qn} \times \bfA) + 2 (\bfv_{* e} \times \bfB) \cdot \bfB 
	\cr 
	+ 2\bigg[ \frac{\grad p_e}{en} &+& \frac{m_e}{e} \left\{ \pdr_t \bfv_e + \bfw_e \times \bfv_{*e} + \half \grad \bfv_e^2 \right\} \bigg] \cdot \bfB \cr
	&=& - \grad \cdot \left[ \phi \bfB + \bfE_{\rm qn} \times \bfA - \frac{2 \tl h_e \bfB}{e} \right] 
	\cr 
	&+& \frac{2 m_e}{e} \left[ \pdr_t \bfv_e + \bfw_e \times \bfv_{*e} + \frac{\grad \bfv_e^2}{2} \right] \cdot \bfB.
	\eeqs
When electron inertia terms are ignored, we see that $\bfE_{\rm qn} \to \bfE_{\rm Hall}$ and magnetic helicity satisfies the local conservation law (\ref{e:mag-hel-loc-cons-Hall-2fluid}). The regularization enters through the electron `swirl' velocity $\bfv_{* e}$ in (\ref{e:hall-electric-field}).

However, even in the Hall ($m_e \to 0$) limit, we do not have an analogue of a conserved cross helicity $\bfv \cdot \bfB$ of R-MHD. For instance, using the electron velocity equation (\ref{e:reg-ion-elec-mom-quasineutral}) and the homogeneous Maxwell equations we find
	\beqs
	\pdr_t (\bfv_e \cdot \bfB) &=& - \bfv_e \cdot \grad \times \bfE - \grad \cdot (\sig_e \bfB) 
	\cr
	&& + \bfB \cdot \bfv_{*e} \times (\grad \times \bfv_e) - \frac{e}{m_e} \bfE \cdot \bfB.
	\eeqs
Substituting for $\bfE_{\rm qn}$ (\ref{e:E-from-e-mom-eqn}), combining terms and taking $m_e \to 0$, we find that unlike for magnetic helicity, the final offending term is not suppressed by $m_e$.
	\beqs
	(\bfv_e \cdot \bfB)_t &+& \grad \cdot (\bfv_{*e} (\bfv_e \cdot \bfB)) \cr
	&=& \bfB \cdot \left ( \pdr_t \bfv_e + \bfw_e \times \bfv_{*e} + \grad(\bfv_e \cdot\bfv_{*e}) \right).
	\eeqs

\subsection{From R-Hall to 1-fluid R-MHD when $e \to \infty$}

To get the regularized 1-fluid MHD model of Ref.~\cite{govind-sonakshi-thyagaraja-pop} from the above R-Hall 2-fluid model we let $e \to \infty$, holding $\la_i$ and $\la_e$ fixed. The limit $e \to \infty$ is a convenient way of restricting attention to frequencies small compared to the cyclotron $\om_{c,l} = e B/ m_l$ and plasma $\om_{p,l} = \sqrt{{n_l e^2}/{m_l \eps_0}}$ frequencies and to length scales large compared to the Debye lengths $\la_{D,l} = \sqrt{{ k_B T_l \eps_0}/{n_l e^2}}$,  gyroradii $r_l = v_{th,l}/\om_{c,l} = \sqrt{k_B T_l m_l}/eB$ and collisionless skin depths $\del_l = c/\om_{p,l} = \sqrt{{m_l}/{\mu_0 n_l e^2}}$.

To switch to one-fluid variables we express $\bfv_i$ and $\bfv_e$ in terms of center of mass velocity $\bfv = (m_i \bfv_i + m_e \bfv_e)/m$ and $\bfj_{\rm flow} = e n (\bfv_i - \bfv_e)$
	\beq
	\bfv_{i,e} = \bfv \pm \frac{m_{e,i}}{m} \frac{\bfj_{\rm flow}}{e n}.
	\label{e:v_i-v_e-to_v-j}
	\eeq
Here $m = m_i + m_e$. The continuity equation $\pdr_t \rho = -\grad \cdot (\rho \bfv)$ for the total mass density $\rho = nm$ is obtained by taking a mass-weighted average of the continuity equations in (\ref{e:cont-eqn-quasineutral})
	\beq
	\pdr_t ((m_i + m_e )n) =  -\grad \cdot (n m_i v_i + n m_e v_e)
	\eeq
The evolution equation for the center of mass velocity $\bfv$ is similarly  obtained from (\ref{e:reg-ion-elec-mom-quasineutral}),
	\beqs
	\bfv_t &+& \frac{m_i}{m}\bfw_i \times \bfv_{*i} + \frac{m_e}{m}\bfw_e \times \bfv_{*e} = - \ov{n m} \grad (p_i + p_e) 
	\cr &&
	- \frac{1}{2m} \grad (m_i \bfv_{i}^2 + m_e \bfv_{e}^2)  + \frac{e}{m}(\bfv_{*i} - \bfv_{*e}) \times \bfB.
	\eeqs
Neglecting terms of order $m_e/m \ll 1$ and introducing $\bfj_* = e n (\bfv_{*i} - \bfv_{*e})$ and $p = p_i + p_e$ we get 
	\beq
	\pdr_t \bfv + \bfw_i \times \bfv_{*i} = -  \ov{\rho} \grad p - \half \grad \bfv_{i}^2 + \frac{1}{\rho}(\bfj_* \times \bfB).
	\eeq
Next we take the limit $e \to \infty$ in (\ref{e:v_i-v_e-to_v-j}) keeping $\bfj_{\rm flow}$ finite so that $\bfv, \bfv_i$ and $\bfv_e$ are all equal, as are $\bfw, \bfw_i$ and $\bfw_e$. Defining $\la = \la_i $, $\bfv_{*i}$ = $\bfv_* = \bfv + \la^2 \grad \times \bfw$. Thus, we arrive at the velocity equation for one-fluid R-MHD,
	\beq
	\pdr_t \bfv + \bfw \times \bfv_{*} = -  \ov{\rho} \grad p - \half \grad \bfv^2 + \frac{1}{\rho}(\bfj_* \times \bfB).
	\label{e:1-fluid-reg}
	\eeq
However unlike in the 2-fluid model $\bfj_*$ is no longer given by $e n (\bfv_{*i} - \bfv_{*e})$. Instead, it is obtained from Amp\`ere's law $\mu_0 \bfj_* = \grad \times \bfB$. On the other hand, taking the limit $e \to \infty$ in the Hall electric field (\ref{e:hall-electric-field}) the pressure gradient term drops out and we get
	\beq
	\bfE_{\rm 1-fluid} = - \bfv_{*e} \times \bfB = - \bfv_{*} \times \bfB.
	\eeq
This identification of $\bfv_{*e}$ with the 1-fluid swirl velocity $\bfv_*$ requires that $\la_e = \la$. Thus, to get the 1-fluid R-MHD model we need to take $\la_i = \la_e = \la$. Finally, Faraday's law (\ref{e:Maxwell-eqn-quasi-neutral}) becomes $\pdr_t \bfB = \grad \times (\bfv_* \times \bfB)$ implying that the solenoidal $\bfB$ is frozen into $\bfv_* $.

\section{Poisson brackets for regularized compressible two-fluid plasmas}
\label{s:PB-2-fluid}

Poisson brackets for (unregularized) two-fluid plasmas were proposed by Spencer and Kaufman \cite{spencer-kaufman} and Holm and Kuperschmidt \cite{holm-kuperschmidt}. The non-trivial PBs are given by 
	\beqs
	\{v_l^{\al} (x), v_l^{\beta} (y) \} &=& \frac{\eps^{\al \beta \gamma}}{m_l n_l} \left( w_l^{\gamma} + \frac{q_l B^{\gamma}}{m_l}\right) \del( x - y), \cr
	\{\bfv_l(x), n_l(y) \} &=& \{n_l(x), \bfv_l (y) \} = \ov{m_l}\grad_y \del( x - y),
	\cr 
	\{E^{\al} (x), B^{\beta} (y) \} &=& \frac{\eps^{\al \beta \gamma}} {\eps_0} \pdr_{y^{\gamma}}\del( x - y)
	 \quad \text{and} \cr 
	\{v_l^{\al} (x), E^{\beta} (y) \} &=& \frac{q_l}{m_l \eps_0} \del^{\al \beta}\del(x - y).
	\eeqs
Here, $l = i,e$ labels species while $\al, \beta ,\gamma$ label Cartesian components. The velocity PBs for a given species are obtained from the Landau PBs $\{v^{\al}, v^{\beta} \} = \eps^{\al \beta \gamma} w^{\gamma} \del(x -y)/\rho$ of fluid mechanics by replacing $\bfw$ by $\bfw + q \bfB/m$ and $\rho$ by $m n$ for each species. This is reminiscent of the results established in \cite{thyagaraja-mc-clements-2}, already mentioned. Similarly, $\{ \bfv_l, n_l \}$ is obtained from Landau's PB $\{ \bfv(x) , \rho(y) \} = \grad_y \del(x-y)$. The rest of the PBs vanish $\{\bfB(x) , \bfB(y) \} = \{\bfv_l, \bfB \} = \{\bfB , n_l \} = \{\bfE , n_l \} = \{\bfE , \bfE \} = \{n_l , n_{l'} \}  = \{\bfv_i , \bfv_e \} = \{\bfv_e, n_i \} = \{\bfv_i , n_e  \}  = 0$. In particular, unlike in 1-fluid MHD \cite{landau,morrison-greene,govind-sonakshi-thyagaraja-pop}, velocities and $\bfB$ commute. Vorticity behaves in a manner similar to $\bfB$: $\{ \bfw_l, n_{l'} \} = \{ \bfw_l, \bfB \} = 0$; $\{ \bfE, \bfw_l \}$ is similar to $\{ \bfE, \bfB \}$:
	\beq 
	\{ E^\al(x) , w_l^\beta(y) \} = \frac{\eps^{\al \beta \gamma} q_l}{\eps_0 m_l} \pdr_{y^{\gamma}}\del(x - y).
	\eeq
Our twirl regularization is natural in the sense that the regularized equations follow from these PBs with the swirl energy (\ref{e:energy-density-2-fluid}) as Hamiltonian. We sketch how this happens. It follows from the PBs that only the kinetic energies contribute to the continuity equations,
	\beqs
	\pdr_t n_l(x) &=& \{ n_l, KE_l \} = \int m_l n_l \bfv_l \cdot \{ n_l(x) , \bfv_l(y) \} dy
	\cr
	&=& \int n_l \bfv_l \cdot \grad_y \del(x-y) = - \grad \cdot (n_l \bfv_l).
	\eeqs
To obtain the velocity equations we note that the following relations hold for the electric (EE), kinetic (KE$_l$), compressional (PE$_l$) and vortical (VE$_l$) energies:
	\beqs
	\{ \bfv_l(x) , {\rm EE} \} &=& \eps_0 \int E^\beta (y) \{ \bfv_l(x) , E^\beta(y) \} dy = \frac{q_l}{m_l} \bfE, \cr
	 \{ \bfv_l(x) , {\rm PE}_l \} &=& \int U'_l \{ \bfv_l(x) , \rho_l (y) \} dy = - \grad U'_l = -\grad h_l,
	\cr
	\{ \bfv_l(x) , {\rm KE}_l \} &=&  \int \left( \rho_l v_l^\beta(y) \{ \bfv_l (x), v_l^\beta(y) \} \right. 
	\cr
	&& \left. \qquad + \frac{\bfv_l^2}{2} \{ \bfv_l (x), \rho_l(y) \} \right) dy 
	\cr
	&=& \bfv_l \times  \left( \bfw_l + \frac{q_l \bfB}{m_l} \right) - \half \grad \bfv_l^2 \;\; \text{and}
	 \cr
	\{ v_l^\al(x) , {\rm VE}_l \} &=& \la_l^2 \rho_l  \int w_l^\beta(y) \eps_{\beta \g \del } \pdr_{y^\g} \{ v_l^\al (x), v_l^\del (y) \} dy 
	\cr
	&=& - \eps_{\al \eta \del} \la_l^2 \left( w_l^{\eta} + \frac{q_l B^{\eta}}{m_l}\right) \eps_{\del \g \beta} \pdr_\g  w_l^\beta 
	\cr
	&=& - \la_l^2 \left[ \left( \bfw_l + \frac{q_l \bfB}{m_l}\right) \times  (\grad \times  \bfw_l) \right]^\al.
	\eeqs
Thus, using $\sig_l = h_l + \half \bfv_l^2$, we get the velocity equations (\ref{e:reg-mom-eq-barotropic-Tw-TB}) for $l=i,e$. If $\{ \bfv_i, n_e \} \ne 0$, the electron pressure would contribute to the ion velocity equation. Faraday's law receives a contribution only from the electric energy:
	\beq
	\pdr_t \bfB(x) 
	=  \eps_0 \int \bfE(y) \cdot \{ \bfB(x), \bfE(y) \} \: dy = - \grad \times \bfE.
	\eeq
Only KE, VE and magnetic energy (ME) contribute to Amp\`ere's law:
	\beqs
	\{ \bfE(x), {\rm KE}_l \} &=& m_l \int n_l v_l^\al \{ \bfE(x), v^\al_l(y) \} dy
	= -\frac{\bfj_{{\rm flow},l}}{\eps_0}, \cr
	\{ \bfE(x), {\rm VE}_l \} &=& \la_l^2 n_l m_l \int w_l^\al \{ \bfE(x), w^\al_l(y) \} dy 
	\cr
	&=& - \frac{\la_l^2 n_l q_l}{\eps_0} (\grad \times \bfw_l)
	= -\frac{\bfj_{{\rm twirl},l}}{\eps_0} \;\; \text{and}
	\cr
	\{ \bfE(x), {\rm ME} \} &=& \int \frac{B^\al}{\mu_0} \{ \bfE(x), B^\al(y) \} dy = \frac{\grad \times \bfB}{\mu_0 \eps_0}.
	\label{e:E-fld-PBs-2-fluid}
	\eeqs
Combining, we see that the swirl current $\bfj_*$ in Amp\`ere's law is the sum of flow and twirl currents:
	\beq
	\pdr_t \bfE = - \ov{\eps_0} \sum_{l} \left( \bfj_{{\rm flow},l} + \bfj_{{\rm twirl},l} \right) + \ov{\mu_0 \eps_0} \grad \times \bfB.
	\label{e:ampere-law-from-pb-reg-2-fluid}
	\eeq

\section{Regularization of $\grad \times \bfB$ in single and two-fluid models}
\label{s:reg-field-curl-PBs-Hamiltonian}

The twirl terms $\bfw_l \times (\grad \times \bfw_l)$ and $\bfB \times (\grad \times \bfw_l)$ in the EOM and the corresponding vortical energies $\half \la_l^2 n_l m_l \bfw_l^2$ can smooth out large velocity gradients and regularize vortical singularities. Similarly, we would like to identify appropriate terms in the EOM to regularize magnetic field gradients and current sheets. Recall from \S \ref{s:energy-cons-2-fluid-incompress} that in the quasi-neutral incompressible case the term $(\la^2/2\mu_0) (\grad \times \bfB)^2$ automatically arose in the conserved energy if the current in Amp\`ere's law is chosen to be the flow current $\bfj_{\rm flow}$ and $\la_i = \la_e = \la$. This approach however does not generalize to compressible flow. In the compressible case, the current in Amp\`ere's law must be the swirl current $\bfj_*$ to guarantee energy conservation. On the other hand, the Poisson bracket formulation gives us a natural way of introducing field gradient energies in compressible flow. Adding the simplest possible positive definite magnetic gradient energy (MGE) term $\int \la_B^2 (\grad \times \bfB)^2/{2 \mu_0} \, d\bfr$ to the Hamiltonian of the single and 2-fluid models and using the relevant PBs to obtain the EOM, we ensure the $L^2$ boundedness of $\grad \times \bfB$.


\subsection{Regularization of $\grad \times \bfB$ in R-MHD}
\label{s:1-fluid-curl-B-reg}

We augment the R-MHD Hamiltonian with a magnetic gradient energy taking $\la_B$ to be a constant cut-off length
	\beq
	H = \int \left[ \frac{\rho \bfv^2}{2}  + U + \frac{\la^2 \rho \bfw^2}{2}  + \frac{\bfB^2}{2 \mu_0} + \frac{\la_B^2}{2 \mu_0} (\grad \times \bfB)^2 \right] \: d\bfr.
	\eeq
Using the non-trivial 1-fluid PBs \cite{landau,morrison-greene}, \; $\{ \rho(x), \bfv(y) \} = \grad_\bfy \del(x-y)$,
	\beqs
	&& \{ v_\al(x), v_\beta(y) \} =\frac{\eps_{\al \beta \g} w_\g}{\rho} \del(x-y) \quad \text{and}  
	\cr
	&& \{ v_\al(x) , B_\beta(y) \} = \frac{\eps_{\al \g \sig} \eps_{\beta \eta \sig}}{\rho(x)} B_\g(x) \pdr_{x^\eta} \del(x-y),
	\eeqs
the continuity and Faraday equations are unchanged
	\beq
 	\pdr_t \rho + \grad \cdot (\rho \bfv) = 0 \quad \text{and} \quad 
	\pdr_t \bfB = \grad \times (\bfv_* \times \bfB).
	\eeq
On the other hand, the velocity equation is modified by
	\beqs
	 && \{v_{\al} (x), {\rm MGE} \} = \frac{\la_B^2}{2 \mu_0}  \int \{ v_{\al} (x), (\grad \times \bfB)^2  \} dy 
	 \cr
	 &=& \frac{\la_B^2}{\mu_0 \rho} \: \eps_{jkl} \eps_{\al mn}\eps_{lpn} B_m \pdr_{x^p} \int \left[(\grad \times \bfB)_j \pdr_{y^k} \del(x - y) \right] dy \cr
	 &=& - \frac{\la_B^2}{\rho \mu_0} \left[\bfB \times \left( \grad \times \left( \grad \times (\grad \times \bfB) \right) \right) \right]_\al.
	\eeqs
Combining this with contributions from  kinetic, potential, vortical and magnetic energies, the velocity equation takes the same form as (\ref{e:1-fluid-reg}) with $\bfj_*$ replaced by the regularized `magnetic swirl' current 
	\beqs
	\mu_0 \bfj_{**} &=& \grad \times \bfB + \la_B^2 \grad \times \left( \grad \times \left(\grad \times \bfB \right) \right) \cr
	 &=& (1 - \la_B^2 \grad^2) (\grad \times \bfB).
	\label{e:magnetic-swirl-current}
	\eeqs
Evidently, $\mu_0 \bfj_{**}$ is the magnetic analogue of $\bfv_* = \bfv + \la^2 \grad \times (\grad \times \bfv)$. Furthermore, $\grad \times \bfB$ is a smoothed version of the regularized current obtained through the application of the integral operator $(1 - \la_B^2 \grad^2)^{-1}$:
	\beq
	\grad \times \bfB = \mu_0 (1 - \la_B^2 \grad^2)^{-1} \, \bfj_{**}.
	\eeq
A similar smoothing operator appears in the {\it non-local} Euler-$\alpha$ equations \cite{Euler-alpha}. As noted in the introduction, these additional terms in the velocity and Faraday equations are quite different from those that appear in XMHD\cite{kimura-morrison,abdelhamid-kawazura-yoshida}. The latter involves the introduction of a $\bfB^* = \bfB + d_e^2 \grad \times ((\grad \times \bfB)/\rho)$  where $d_e$ is a constant normalized electron skin depth, rather than a swirl current $\bfj_{**}$. For instance, this leads to a new term $\bfj \times \bfB^*$ in both the velocity equation and in the electric field in XMHD.

\subsection{Regularization of field curl in the two-fluid model}

As for the single fluid, we augment the 2-fluid Hamiltonian (\ref{e:energy-density-2-fluid}) with a magnetic gradient energy:
	\beqs
	H &=& \int \left[ \sum_l \left( \half m_l n_l \left( \bfv_l^2  + \la_l^2 \bfw_l^2 \right) + U_l(\rho_l) \right) \right.
	\cr
	&& \left. + \frac{\bfB^2}{2 \mu_0} + \frac{\eps_0 \bfE^2}{2} + \ov{2 \mu_0} \la_B^2 (\grad \times \bfB)^2 \right] \: d\bfr.
	\eeqs
Like before, $\la_B$ is a constant cut-off length. Using the 2-fluid PBs of \S \ref{s:PB-2-fluid}, we see that the momentum, continuity and Faraday equations remain unchanged since $\bfv_i, \bfv_e, n_i, n_e$ and $\bfB$ commute with the magnetic field. We do not introduce a $(\grad \times \bfE)^2$ term in $H$ as it would modify Faraday's law. The evolution equation for the electric field is modified by the term:
	\beqs
	\left\{ \bfE(x), {\rm MGE} \right\}
	= \frac{\la_B^2}{\mu_0 \eps_0}\grad \times (\grad \times (\grad \times \bfB))
	= - \frac{\bfj_B}{\eps_0}. \quad \;
	\label{e:j_B-defn-2-fluid}
	\eeqs
Combining with (\ref{e:E-fld-PBs-2-fluid}), Amp\`ere's law (\ref{e:ampere-law-from-pb-reg-2-fluid}) becomes
	\beq
	\mu_0 \eps_0 \pdr_t \bfE = \grad \times \bfB - \mu_0 \bfj_* - \mu_0 \bfj_B.
	\eeq
Here, $\bfj_* = \bfj_{\rm flow} + \bfj_{\rm twirl}$. Now, we can define a new current density $\bfj_{**} = \bfj_* + \bfj_B$. Note that (\ref{e:j_B-defn-2-fluid}) implies $\grad \cdot \bfj_B = 0$. Thus $\bfj_B$ and $\bfj_{\rm twirl}$ are like magnetization currents in material media/plasmas. We notice that the introduction of the MGE in the Hamiltonian has apparently very different effects in the single and two-fluid models. In the former, the velocity equation is modified while it is the Amp\`ere equation that is modified in the latter. However, the two are closely related. In fact, upon taking the limits $\eps_0 \to 0,m_e \to 0 $ and $e \to \infty$, the 2-fluid current density $\bfj_{**}$ exactly matches the magnetic swirl current (\ref{e:magnetic-swirl-current}) appearing in the Lorentz force term of the single fluid velocity equation.

\section{Discussion}
\label{s:discussion}

In this paper, we have extended the conservative twirl regularization of our earlier work \cite{thyagaraja,govind-sonakshi-thyagaraja-pop,govind-sonakshi-thyagaraja-arXiv} to dissipationless compressible two-fluid plasmas. This involves vortical and magnetic twirl terms $\lambda_l^2 (\bfw_l + \frac{q_l}{m_l} \bfB) \times (\grad \times \bfw_l)$ in the velocity equations for ions and electrons. We find that $\la_l^2 n_l$ must be constant for energy conservation, so that $\la_l$ behaves likes $\la_{D}$ or $c/\om_{p,l}$. The key difference between the regularized and unregularized two-fluid models is that the flow current $\bfj_{\rm flow} = \sum_l q_l n_l \bfv_l$ in Amp\`ere's law is augmented by a solenoidal `twirl' current $\sum_l \grad \times (\grad \times \la_l^2 \bfj_{{\rm flow},l})$ analogous to magnetization currents in material media. This leads to locally conserved momenta and a positive definite swirl energy $E^*$. In addition to kinetic, compressional and electromagnetic contributions, $E^*$ includes a vortical energy density $\sum_l \la_l^2 n_l m_l \bfw_l^2$, thus placing an \`a priori upper bound on the enstrophy of each species.  It is noteworthy that our twirl-regularized two-fluid equations follow from the Hamiltonian $E^*$ using unchanged the Poisson brackets of \cite{spencer-kaufman,holm-kuperschmidt}. This PB formalism shows that among regularizations preserving the continuity equations and symmetries of the ideal system, our twirl regularization terms are unique and minimal in non-linearity and space derivatives of velocities. It is also employed to regularize magnetic field curls in the compressible models by adding $(\la_B^2/2\mu_0) \int (\grad \times \bfB)^2 \: d\bfr$ to $E^*$ so that field and velocity curls are $L^2$-bounded. By taking suitable successive limits we get a hierarchy of compressible and incompressible regularized plasma models (quasi-neutral two-fluid, Hall and 1-fluid MHD). Interestingly, in the incompressible two-fluid case alone, it is also possible to choose the current as $\bfj_{\rm flow}$, which leads to a conserved swirl energy that automatically includes a $(\la^2/2\mu_0) \int (\grad \times \bfB)^2 \: d\bfr$ term in $E^*$. Furthermore, the assumption of local short-distance cut-offs $\la_{i,e}$ limits the number of effective degrees of freedom, thus considerably extending results on the CHM model\cite{davies-mccarthy-thyagaraja} to the full 3-D two-fluid equations. This feature is crucial to numerical modeling of conservative plasma dynamics and consequently provides a viable  framework to investigate statistical theories of turbulence in these systems. 
While we have regularized vortical and field singularities, there remains the question of conservatively regularizing density/pressure gradients in shocks. This requires additional terms\cite{govind-sonakshi-thyagaraja-pop} in the Hamiltonian which could alter the continuity and energy equations analogous to the KdV-type regularization of the kinematic wave equation in one-dimension.

A natural question concerns the effect of our twirl regularization in specific fluid and plasma systems of interest. We have examined this in a few representative steady flows\cite{govind-sonakshi-thyagaraja-pop,govind-sonakshi-thyagaraja-arXiv}: a rotating columnar vortex and its extension to MHD, a vortex sheet, compressible plane flow, channel flow and variants of Hill's vortex. In all these steady flows, the non-linear regularized equations are under-determined as in ideal Euler or ideal MHD. For instance, in our rotating columnar vortex model for a tornado\cite{govind-sonakshi-thyagaraja-pop} with core radius $a$, the equations determine the density if the vorticity distribution is prescribed. In a layer whose width can be of order the regularization length $\lambda \ll a$, the vorticity smoothly drops from its value in the core to that in the periphery. We find that the regularization relates this decrease in vorticity to a rise in density. On the other hand, vorticity is allowed to have an unrestricted jump across the layer in the unregularized model while $\rho$ is continuous and its increase is unrelated to the drop in vorticity. Similarly, the regularization can smooth the vorticity in a magnetized columnar vortex\cite{govind-sonakshi-thyagaraja-pop}. Given vorticity and current profiles, the density profile is determined. While the Lorentz force tends to pinch the column, the twirl force points outwards for radially decreasing vorticity. An analogue of Hill's vortex, a cylindrical vortex in pipe-like flow was considered in Reference~\cite{govind-sonakshi-thyagaraja-arXiv}. The flow is irrotational outside an infinite circular cylinder of radius $a$ with vorticity purely azimuthal inside the cylinder. The regularized equations with appropriate BCs were solved numerically and unlike in the unregularized case, the vorticity was found to be continuous across $r = a$. In modeling a vortex sheet\cite{govind-sonakshi-thyagaraja-pop}, we found steady solutions to the regularized equations that smooth discontinuous changes in vorticity over a layer of thickness $\gtrsim \lambda$. A regularized analogue of a Bernoulli-like equation implies a reduction in density on the sheet compared to its asymptotic values: depending on the relative flow Mach number, the decrease can be significant when the thickness of the sheet is comparable to the regulator $\lambda$. These examples show that twirl-regularized steady flows can be more regular than the corresponding ideal ones. They also serve as a starting point for numerical simulations of time-dependent flows. An interesting example that is currently under investigation concerns the effect of our regularizations on the growth of perturbations to vortex/current sheets and their non-linear saturation. A problem of fundamental importance is the initial value problem in 3D, say with periodic BCs. We would like to numerically simulate the regularized equations and determine the spectral distribution of energy and enstrophy over long times.

\begin{acknowledgments} 

This work was supported in part by the Infosys Foundation. AT thanks Prof. Mark Birkinshaw for useful discussions and CMI for hospitality and support. We would also like to thank a referee for suggesting improvements to the paper.

\end{acknowledgments}

\appendix

\section{Minimality of twirl regularization in Hamiltonian formulation}
\label{s:minimality}

Here we address the question of minimality/uniqueness of the twirl regularization, firstly in the context of neutral flows. We show that the twirl term $\la^2 \bfw \times (\grad \times \bfw)$ is the minimal symmetry-preserving conservative regularization term that can be added to the Euler equation while retaining the usual continuity equation and standard Hamiltonian formulation. The Euler equation is invariant under space-time translations, rotations, time reversal $T$ and parity $P$. We seek regularization term(s) involving $\rho$, $\bfv$ and derivatives of $\bfv$ that may be added to the Euler equation while preserving these symmetries. Any such term must be even under $T$, odd under $P$, not involve either $\bfr$ or $t$ explicitly, and transform as a vector under rotations. Furthermore, we seek terms with as few spatial derivatives, no time derivatives and as low a non-linearity in $\bfv$ as possible. The term must preferably involve a (possibly dynamical) length $\la$ that can play the role of a short-distance cut-off. However, there are very many such terms even if we restrict to those quadratic in $\bfv$ with at most three derivatives [E.g. $\la^2 \bfw \times (\grad \times \bfw), \la^2 (\bfw \cdot \grad) \bfw \;\text{or}\; \la^2 \eps^{ijk} \pdr_j w_l \pdr_l v_k$] and it is an arduous task to identify all of them. We may simplify our task by requiring that the regularized equations follow from a Hamiltonian and the standard Landau PBs. Thus we seek a positive definite regularization term ${\cal H}_R$ involving $\bfv$ and its derivatives (dependence on $\rho$ is then fixed by dimensional arguments) that may be added to the ideal Hamiltonian density ${\cal H}_I = (1/2) \rho v^2 + U(\rho)$. The possibility of including derivatives of $\rho$ in ${\cal H}_R$ will be considered elsewhere. The advantage of working with the Hamiltonian is that we need only consider scalars rather than the more numerous vectors [regularizations that do not admit a Hamiltonian-PB formulation would however not be identified by this approach]. Due to the PB structure $(\{\bfv , \bfv \} \propto \pdr \bfv)$, the number of spatial derivatives in the velocity equation $\bfv_t = \{ \bfv, H \}$ is one more than that in $H$ and the degree of non-linearity in $\bfv$ is the same as in $H$. Thus, ${\cal H}_R(v_i, \pdr_j v_i, \ldots)$ must be a $P$ and $T$-invariant scalar with a minimal number of derivatives and minimal non-linearity in $\bfv$. It would be natural to ask that ${\cal H}_R$ be non-trivial in the incompressible limit, so that it may regularize vortical singularities in such flows. However, we find that such a restriction is not necessary. On the other hand, we do require that the regularization leave the continuity eqaution $\rho_t = \{\rho, H\}= -\grad \cdot (\rho \bfv)$ unaltered i.e., $\{ \rho, H_R\} = 0$, assuming decaying or periodic boundary conditions (BCs) in a box. Now, for ${\cal H}_R$ to be $P$-even, the sum of the number of spatial derivatives and degree of non-linearity in $\bfv$ must be even. $T$-invariance as well as positive definiteness require that the degree of ${\cal H}_R$ in $\bfv$ be even. Thus we begin by listing all scalars at most quadratic in $\bfv$ with at most two derivatives. They are obtained by picking coefficient tensors $C^{ijk\ldots}$ below as linear combinations of products of the rotation-invariant tensors $\del^{ij}$ and $\eps^{ijk}$:
	\beqs
	1v,1\pdr: \: &&  C^{ij} \pdr_i v_j = \del^{ij} \pdr_i v_j  = \grad \cdot \bfv, \cr
	1v,2\pdr: \: &&  C^{ijk} \pdr_i \pdr_j v_k = \eps^{ijk}\pdr_i \pdr_j v_k = 0, \cr
	2v,0\pdr: \: && C^{ij} v_i v_j = \del^{ij}v_i v_j =\bfv^2, \cr
	2v,1\pdr: \: && C^{ijk} v_i \pdr_j v_k = \bfv \cdot \bfw; C^{ijk} \pdr_i (v_j v_k) = 0.
		\eeqs
$T$-invariance eliminates $\grad \cdot \bfv$, $P$-invariance eliminates $\bfv \cdot \bfw$ while $\bfv^2$ is already present in ${\cal H}_I$. Thus we are left with quadratic scalars with two derivatives:
	\beqs
	 C^{ijkl} v_i \pdr_j \pdr_k v_l &=& (c_1 + c_3) \bfv \cdot \grad (\grad \cdot \bfv) + c_2 \bfv \cdot \grad^2 \bfv \cr
	 C^{ijkl} \pdr_i v_j \: \pdr_k v_l &=& c_4 (\pdr_i v_j)^2 + c_5 \pdr_i v_j \: \pdr_j v_i + c_6 (\grad \cdot \bfv)^2 
	\cr
	C^{ijkl} \pdr_i \pdr_j (v_k v_l) &=& c_7 \grad^2 \bfv^2 + (c_8 + c_9) (2 \bfv \cdot \grad (\grad \cdot \bfv)  \cr
	&& + (c_8 + c_9)((\grad \cdot \bfv)^2 + \pdr_i v_j \: \pdr_j v_i).
	\eeqs
Here, $C^{ijkl}$ has been written as a linear combination of the products $\del^{ij} \del^{kl}$, $\del^{il} \del^{jk}$ and $ \del^{ik} \del^{jl}$. Note that the order of indices in $C^{ijk\cdots}$ does not matter: E.g., the space of scalars spanned by $C^{ijkl} \pdr_i \pdr_j (v_k v_l)$ and $C^{ljki} \pdr_i \pdr_j (v_k v_l)$ are the same. The coefficients in the linear combination must be functions of $\rho$ alone and on dimensional grounds must be constants $c_n = \la_n^2 \rho$ where $\la_n$ are position-dependent short-distance cutoffs. The identity $\grad^2 \bfv^2 = 2 \bfv \cdot \grad^2 \bfv + 2 (\pdr_i v_j)^2$ implies there are only five such linearly independent scalars. Since enstrophy density $\bfw^2 = (\pdr_i v_j)^2 - (\pdr_i v_j)(\pdr_j v_i)$ is a physically interesting linear combination, it is convenient to choose the basis for such scalars as $S_1 = \bfw^2$, $S_2 = \bfv \cdot \grad^2 \bfv$, $S_3 = (\pdr_i v_j)(\pdr_j v_i)$, $S_4 = (\grad \cdot \bfv)^2$ and $S_5 = \bfv \cdot \grad (\grad \cdot \bfv)$. We will now argue that $\bfw^2$ is the only independent regularizing term. Consider first the incompressible case where $S_4 = S_5 = 0$. Integrating by parts, $\int S_3 d\bfr = 0$ for decaying/periodic BCs. Furthermore, $\int S_2 \: d\bfr = \int \bfv \cdot \left[\grad (\grad \cdot \bfv) - \grad \times \bfw \right] d\bfr = \int \bfw^2 d\bfr$. Thus for incompressible flow we have shown that $\la^2 \rho \bfw^2$ is the only independent, positive definite $(\la^2 \rho > 0)$, Galilean-invariant regularization term. For compressible flow, we will not consider regularizations that alter the continuity equation, leaving that possibility for the future. Thus we require $\{ \rho, H_R \} = 0$. Since $\{\rho, \bfw\} = 0$, the term  $\bfw^2$ will not affect the continuity equation. On the other hand, the four other possibilities do modify it:
	\beq
	\{ \rho, \int (S_3 \;,\; S_4 \;, \; -S_2 \;, \; -S_5 ) \: d\bfr \} 
	= 2 \grad^2 (\grad \cdot \bfv).
	\eeq
To preserve the continuity equation, we may consider sums or differences of the above terms. Thus we replace the $S_{1, \cdots, 5}$ basis with the new basis $\tl S_1 = \bfw^2$, $\tl S_2 = \bfv \cdot \grad^2 \bfv + (\pdr_i v_j)(\pdr_j v_i)$, $\tl S_3 = \bfv \cdot \grad^2 \bfv + (\grad \cdot \bfv)^2$, $\tl S_4 = \bfv \cdot \grad (\grad \cdot \bfv) + (\pdr_i v_j)(\pdr_j v_i)$ and $\tl S_5 = \bfv \cdot \grad (\grad \cdot \bfv) + (\grad \cdot \bfv)^2$. As before, $\int \tl S_2 \: d\bfr =  \int \tl S_3 \: d\bfr = -\int \bfw^2 \: d\bfr$ and $\int \tl S_4 \: d\bfr = \int \tl S_5 \: d\bfr = 0$. Subject to these BCs, we have shown that $H_R = \int \la^2 \rho \bfw^2 \: d\bfr$ is the only positive-definite velocity-dependent regularizing term in the Hamiltonain that (a) preserves parity, time-reversal, translation, rotation and boost symmetries of the system, (b) does not alter the continuity equation and (c) involves at most two spatial derivatives and is at most quadratic in $\bfv$. We conclude that with the standard PBs, the twirl term $- \la^2 \bfw \times (\grad \times \bfw)$ with the constitutive relation $\la^2 \rho = $ const., is the only possible regularizing term in the Euler equation that is at most quadratic in $\bfv$ with at most 3 derivatives while possessing properties (a) and (b).

Extending these arguments to 2-fluid plasmas, we may add a linear combination of $\bfw_i^2$, $\bfw_e^2$ and $\bfw_i \cdot \bfw_e$ to the Hamiltonian density. The cross term $\bfw_i \cdot \bfw_e$ leads to {\it direct} interspecies interaction in the velocity equations which we wish to avoid, preferring the ions and electrons to interact via the electromagnetic field. Thus we are left with $\bfw_i^2$ and $\bfw_e^2$ which lead to the vortical energies of ions and electrons considered in \S \ref{s:reg-eqns-2-fluid-compress}.



\end{document}